\documentclass[showpacs]{revtex4}
\usepackage{amsfonts}
\usepackage{amsmath}
\usepackage{amssymb}
\usepackage{graphicx}

\begin{document}

\title{Optical purification of materials based on atom walking in traveling-wave lights}
\author{Wenxi Lai}\email{wxlai@pku.edu.cn}
\affiliation{School of Applied Science, Beijing Information Science and Technology University, Beijing 100192, China}

\begin{abstract}
\textbf{Abstract}$-$An optical method for precise purification of chemical elements is introduced in this paper. The materials are supposed to be in the states of gaseous beams, which are coherently coupled to an external traveling light during purification. Before decoherence occurs, atoms periodically move in the light with different speeds that depends on masses and optical transition wave lengths of these atoms. The speed gradient leads to deflections of different atoms in different directions. The model is described by Schr\"{o}dinger equations with analytical results. This method could be used for some hardly separable atoms and isotopes depending on the condition of atom coherent time. The present work opens a platform for applications of cold atom technology in the purification of atoms and molecules.
\end{abstract}


\maketitle
Natural materials are always mixed with foreign matters which affect their physical or chemical properties. Material purification is the physical separation of a chemical substance of interest from foreign or contaminating substances. It plays important role in many fields, such as medicine, chemistry, manufacturing and micro-fabrication. Especially the material challenges are limiting urgent progress in frontier technology, for example, quantum computing hardware platforms~\cite{Leon}, single photon sources~\cite{Senichev}, high-temperature superconductivity~\cite{Keimer,Paglione}, semiconductor quantum dots~\cite{Arquer} and so on. There are many well developed methods of purifications for different situations~\cite{Haken}. Electromagnetic separation depends on charged particles, ionized atoms for example. In the scheme of diffusion, atoms with different masses diffuse with different velocities. In centrifugation, heavier atoms need larger centrifugal forces. Evaporation or distillation makes use of the boiling point that commonly related to the particle masses. In the process of electrolysis, decomposition difficulties of molecules are different for isotopes that construct the molecules.

\begin{figure}
  \includegraphics[width=8cm]{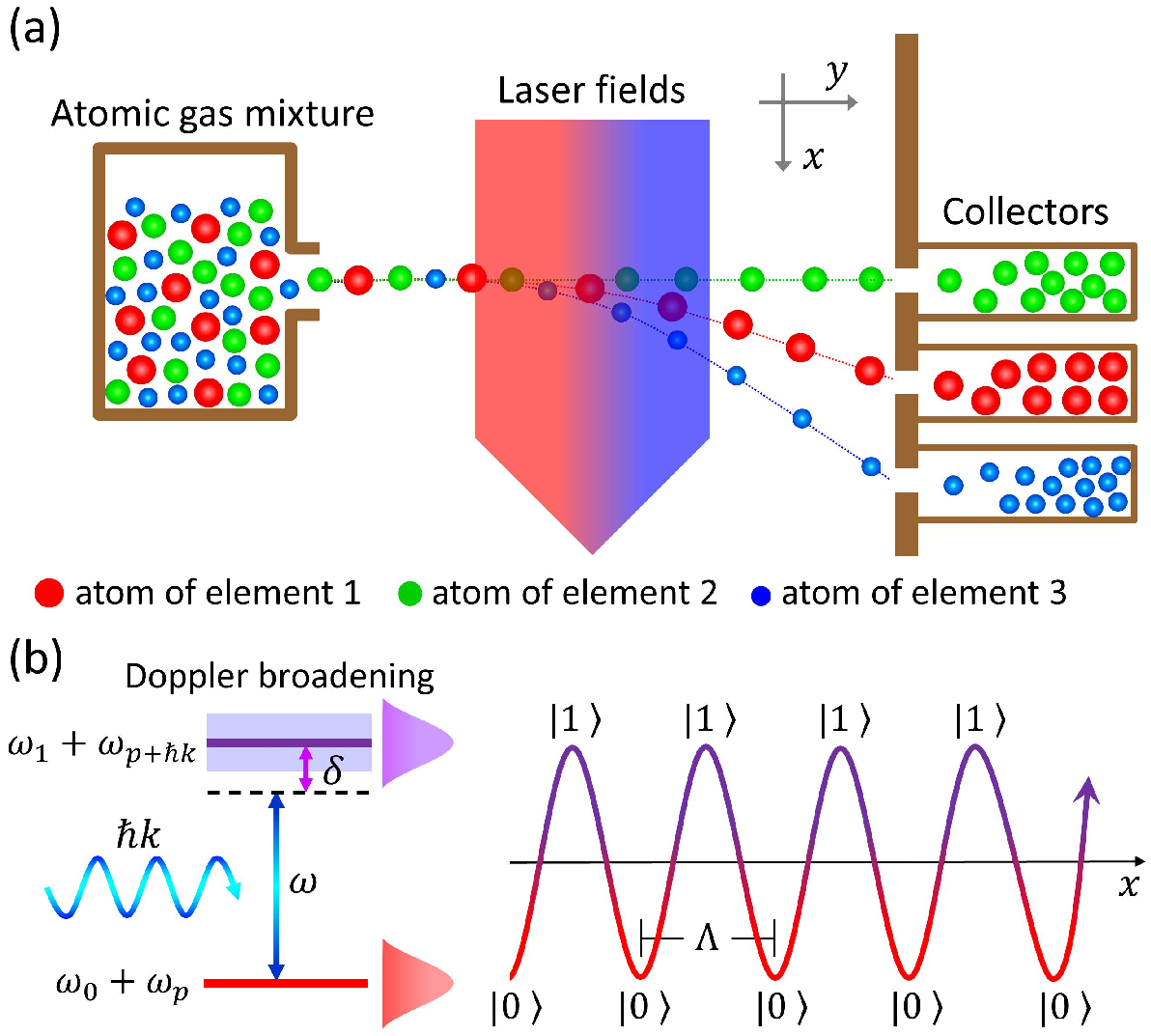}\\
  \caption{(Color on line) (a) Schematic illustration of separation and purification. Three kinds of atoms are considered in this picture. (b) Energy structure and electromagnetically induced walking of the atom. $\omega$ is frequency of the electromagnetic field and $\Lambda$ is one step length of the atom walk.}\label{fig1}
\end{figure}

This paper proposes an optical method of purification based on coherent motion of atoms in traveling lights. The idea is originated from walking of cold atoms in a traveling-wave light with speed depending on atom mass and Rabi frequency~\cite{Lai}. It has been long time that momentum exchange between atom and photon interactions was exploited~\cite{Cook,Dalibard}. This effects have been applied in atom cooling~\cite{Phillips}, atom trapping~\cite{Doherty} and atom acceleration~\cite{Eichmann}. Now, we take advantage of this mechanical effect of atom-light interaction to purify materials. The model considered here is a dilute gaseous atomic beam passing through a monochromatic coherent light. Atomic and molecular beams with low temperature could be realized in experiment~\cite{Maxwell}. Ultra-cooled atom beams for clock~\cite{Elgin} and inertial sensing~\cite{Kwolek} has been reported recently. In our present configuration, atom beam with different chemical elements or isotopes in optical fields would moves like the behaviors in Stern-Gerlach experiment as shown in Fig.~\ref{fig1} (a).

Here, we mainly consider diluted cold neutral atoms where atom-atom interactions could be neglected~\cite{Gattobigio,Couvert}. It may allow us to describe this process with the interaction between any individual atom and an external light field as sketched in Fig.~\ref{fig1} (b). In this model, atoms are not  distinguished as Bosons or Fermions. Hamiltonian of the atom under a traveling light $\vec{\textbf{E}}(\textbf{x},t)=\vec{e_{z}}E_{0}cos(\omega t-k\textbf{x})$ could be described in momentum space as
\begin{eqnarray}
&&\textbf{H}=\sum_{n=0,1}\hbar\omega_{n}|n\rangle\langle n|+\int d p \frac{p^{2}}{2M}|p\rangle\langle p|-\frac{\hbar \Omega}{2}\int d p (e^{-i\omega t}|1,p+\hbar k\rangle\langle 0,p|+e^{i\omega t}|0,p-\hbar k\rangle\langle 1,p|),
\label{eq:Hamiltonian1}
\end{eqnarray}
where $\hbar\omega_{0}$ and $\hbar\omega_{1}$ represent energy of atom two internal states. $\frac{p^{2}}{2M}$ is kinetic energy of the atom with the atom center of mass $M$. $\Omega=|\mu|E_{0}/\hbar$ is the Rabi frequency with dipole moment of the atom $\mu=\langle 0|e\textbf{z}|1\rangle$~\cite{Scully}. Here, $\Omega$ would be taken as real for a definite phase of the dipole moment and light. The action of translation operator, $e^{\pm ik\textbf{x}}|p\rangle=|p\pm\hbar k\rangle$, has been used in derivation of the above Hamiltonian. Evolution of the wave function $|\Psi(t)\rangle$ of the atom could be solved through the Schr\"{o}dinger equation $(i\hbar \frac{\partial}{\partial t}-\textbf{H})|\Psi(t)\rangle=0$. It would be convenient for one to take the unitary operator $e^{i\textbf{H}_{0}t/\hbar}$ with $\textbf{H}_{0}=\hbar\omega_{0}|0\rangle\langle 0|+(\hbar\omega_{0}+\hbar \omega)|1\rangle\langle 1|$ to achieve the Schr\"{o}dinger equation in interaction picture $(i\hbar \frac{\partial}{\partial t}-\textbf{V})|\varphi(t)\rangle=0$, where the wave function can be expanded with probability distribution functions as $|\varphi(t)\rangle=\int d p \sum_{n}\varphi_{n}(p,t)|n,p\rangle$. In a subspace composed of the basic states $\{|0,p\rangle\}$, $|1,p+\hbar k\rangle\}$, the equation of motion have the matrix form
\begin{eqnarray}
[i\frac{\partial}{\partial t}-\left(\begin{array}{cccc}
     \omega_{p} & -\frac{\Omega}{2} \\
    -\frac{\Omega}{2} & \Delta +\omega_{p+\hbar k}
  \end{array}\right)]\left[\begin{array}{cccc}
     \varphi_{0,p}(t) \\
    \varphi_{1,p+\hbar k}(t)
  \end{array}\right]=0,
\label{eq:equ-motion}
\end{eqnarray}
where we defined that $\omega_{p}=\frac{p^{2}}{2M\hbar}$, $\omega_{p+\hbar k}=\frac{(p+\hbar k)^{2}}{2M\hbar}$ and the detuning $\Delta=\omega_{1}-\omega_{0}-\omega$. Eq.\eqref{eq:equ-motion} can be easily diagonalized and the eigenfrequencies are $ W_{0,1}=\frac{1}{2}(\Delta+\omega_{p+\hbar k}+\omega_{p}\mp\Sigma)$, in which the effective Rabi frequency is $\Sigma=\sqrt{\delta^{2}+\Omega^{2}}$. In the energy shift $\delta=\Delta+\frac{p \hbar k}{M}+\frac{\hbar^{2}k^{2}}{2M}$, the second term $\frac{p \hbar k}{M}$ represents Doppler shift and the third term $\frac{\hbar^{2}k^{2}}{2M}$ is back action shift.

\begin{figure}
  \includegraphics[width=8cm]{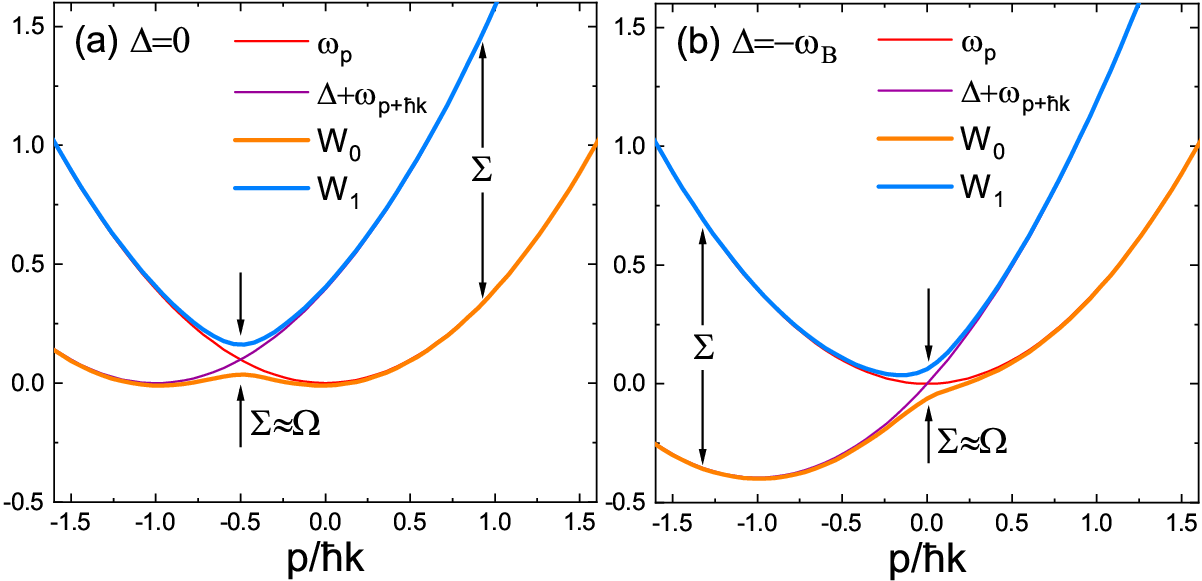}\\
  \caption{(Color on line) Comparisons between the original energy levels and eigenfrequencies of the atom with (a) resonant coupling and (b) a detuning equal to the back action shift.}
  \label{fig2}
\end{figure}

The coherent optical coupling changes the free evolution atom energy structure into a chiral motion of energy as plotted in Fig.~\ref{fig2}. It reveals spin-orbit coupling of the neutral atoms~\cite{Galitski,Lin,Livi}, which is also significant for the realization of topological chiral edge states~\cite{Goldman,Kanungo}. What important for our present task is the strong coupling regime that illustrated in Fig.~\ref{fig2} at resonant coupling and a detuning. In the strong coupling regime $\Omega\gg\delta$, we have $\Sigma\approx\Omega$. In this regime, noise from the Doppler shift and the back action shift in $\delta$ would be removed. Consequently, coherent periodic walking of individual atom could be observed as shown later. The strong coupling regime allows us to write analytical results of Eq.\eqref{eq:equ-motion} as $|\varphi(t)\rangle=\int d p (\varphi_{0}(p,t)|0,p\rangle+\varphi_{1}(p+\hbar k,t)|1,p+\hbar k\rangle)$ with the amplitudes
\begin{eqnarray}
\varphi_{0}(p,t)=(A_{+}e^{i\frac{\Sigma t}{2}}+A_{-}e^{-i\frac{\Sigma t}{2}})e^{-\frac{i}{2}(\Delta+\omega_{p+\hbar k}+\omega_{p})t},
\label{eq:solution1}
\end{eqnarray}
\begin{eqnarray}
\varphi_{1}(p+\hbar k,t)=(B_{+}e^{i\frac{\Sigma t}{2}}+B_{-}e^{-i\frac{\Sigma t}{2}})e^{-\frac{i}{2}(\Delta+\omega_{p+\hbar k}+\omega_{p})t},
\label{eq:solution2}
\end{eqnarray}
where the coefficients that satisfy normalization condition are $A_{\pm}=\frac{1}{2\Sigma}[(\Sigma\pm\delta)\varphi_{0}(p,0)\pm\Omega\varphi_{1}(p+\hbar k,0)]$ and $B_{\pm}=\frac{1}{2\Sigma}[(\Sigma\mp\delta)\varphi_{1}(p+\hbar k,0)\pm\Omega\varphi_{0}(p,0)]$. Initial wave packet is supposed to be a Gaussian function $|\varphi(0)\rangle=\int dp \sum_{n}\varphi_{n}(p,0)|n,p\rangle$, where the probability distribution function is $\varphi_{n}(p,0)=\frac{C_{n}}{\pi^{1/4}\sqrt{\Pi}}e^{-(p-p_{c})^{2}/2\Pi^{2}}$. $C_{n}$ denotes probability amplitude of the atom internal electronic state $|n\rangle$. $p_{c}$ represents the center momentum and $\Pi$ is the characteristic momentum.

The averaged momentum of the atom wave packet can be calculated through the formula $\langle p(t)\rangle=\langle\varphi(t)|p|\varphi(t)\rangle$. From the function of momentum $\langle p(t)\rangle$ we can obtain displacement of the atom $\langle x(t)\rangle=\langle x(0)\rangle+\int_{0}^{t}\frac{\langle p(t)\rangle}{M}dt$. In the strong coupling regime, $\Sigma\approx\Omega$,the displacement of the atom wave packet has the following simple form
\begin{eqnarray}
\langle x(t)\rangle=\langle x(0)\rangle+\frac{p_{c}}{M}t+(C_{0}^{2}-C_{1}^{2})\frac{\hbar k}{2M} (t-\frac{sin(\Omega t)}{\Omega}).
\label{eq:disp}
\end{eqnarray}
Eq.\eqref{eq:disp} describes the coherent walking of single atoms along the $x$ direction. Then, from the formula of atom wave packet displacement, velocity of the atom in the $x$ direction can be achieved as $\langle\upsilon(t)\rangle=\frac{p_{c}}{M}+(C_{0}^{2}-C_{1}^{2})\frac{\hbar k}{2M} (1-cos(\Omega t))$. It is obvious that displacement and velocity of the atom is changing periodically with the time periodicity $T=\frac{2\pi}{\Omega}$. The averaged velocity $\bar{\upsilon}=\frac{1}{T}\int_{0}^{T}\langle\upsilon(t)\rangle dt$ in a periodic time is
\begin{eqnarray}
\bar{\upsilon}=\frac{p_{c}}{M}+(C_{0}^{2}-C_{1}^{2})\frac{\hbar k}{2M}.
\label{eq:aver-v}
\end{eqnarray}
Considering the experimental set up in Fig.~\ref{fig1} (a), initial velocity in $x$ coordinate can be zero, $p_{c}=0$. In addition, the input atoms are assumed to be in the ground state $C_{0}=1$ now. It results the simple relation between atom velocity and its mass-wave length product as $\bar{\upsilon}=\frac{h}{2M\lambda}$, where $h$ is the Plank constant. The formula of $\bar{\upsilon}$ reveals that chemical elements could be separated physically in principle as soon as the product of atom masses $M_{i}$ and a resonant wave lengths $\lambda_{i}$ of corresponding atoms are different in a mixed atomic gas, namely, $M_{i}\lambda_{i}\neq M_{j}\lambda_{j}$ for $i\neq j$.

\begin{table}[hh]
  \caption{Atom samples ($u=1.67\times10^{-27}$ kg)}
  \centering
  \begin{tabular}{c c c c c}
    \hline
    Element & Mass $M$ (u) & Transition $|0\rangle\rightarrow|1\rangle$ & Wavelength $\lambda$ (nm) & Speed $\bar{\upsilon}$ (m/s)\\[0.5ex]
    \hline
    $^{7}$Li & 7.016004 & $2s$ $^{2}$S$_{1/2}-2p$ $^{2}$P$^{0}_{1/2}$ & 670.7926 & 0.042153 \\
    $^{12}$C & 12.000000 & $2s^{2}2p^{2}$ $^{3}$P$_{0}-2s^{2}2p(^{2}$P$^{0})3s$ $^{3}$P$^{0}_{1}$ & 165.6928 & 0.099775\\
    $^{20}$Ne & 19.992435 & $2p^{6}$ $^{1}$S$_{0}-2p^{5}(^{2}$P$^{0}_{1/2})4s$ $^{2}$[1/2]$^{0}$ & 626.8232 & 0.01583 \\
    $^{24}$Mg & 23.985042 & $3s^{2}$ $^{1}$S$_{0}-3s3p$ $^{1}$P$^{0}_{1}$ & 285.21251 & 0.0290 \\
    $^{25}$Mg & 24.985837 & $3s^{2}$ $^{1}$S$_{0}-3s3p$ $^{1}$P$^{0}_{1}$ & 285.21251 & 0.027838 \\
    $^{26}$Mg & 25.982593 & $3s^{2}$ $^{1}$S$_{0}-3s3p$ $^{1}$P$^{0}_{1}$ & 285.21251 & 0.02677 \\
    $^{28}$Si & 27.976927 & $3s^{2}3p^{2}$ $^{3}$P$_{0}-3s^{2}3p4s$ $^{3}$P$^{0}_{1}$ & 251.4316 & 0.028202 \\
    $^{40}$Ca & 39.962591 & $4s^{2}$ $^{1}$S$_{0}-4s4p$ $^{1}$P$^{0}_{1}$ & 422.6727 & 0.011745 \\
    $^{48}$Ti & 47.947947 & $3d^{2}4s^{2}$ a$^{3}$F$_{2}-3d^{2}(^{3}$F$)4s4p(^{3}$P$^{0})$ z$^{3}$D$^{0}_{1}$ & 501.4186 & 0.0082515 \\
    $^{56}$Fe & 55.934939 & $3d^{6}4s^{2}$ a$^{5}$D$_{4}-3d^{6}(^{5}$D$)4s4p(^{3}$P) z$^{5}$P$^{0}_{3}$ & 248.32708 & 0.014282 \\
    $^{59}$Co & 58.933198 & $3d^{7}4s^{2}$ a$^{4}$F$_{9/2}-3d^{7}(^{4}$F$)4s4p(^{3}$P$^{0})$ z$^{4}$F$^{0}_{9/2}$ & 352.6850 & 0.0095446 \\
    $^{69}$Ga & 68.925580 & $4s^{2}4p$ $^{2}$P$^{0}_{1/2}-4s^{2}5s$ $^{2}$S$_{1/2}$ & 403.2984 & 0.0071367 \\
    $^{85}$Rb & 84.911794 & $5s$ $^{2}$S$_{1/2}-5p$ $^{2}$P$^{0}_{3/2}$ & 780.027 & 0.0029952 \\
    $^{87}$Rb & 86.909187 & $5s$ $^{2}$S$_{1/2}-5p$ $^{2}$P$^{0}_{3/2}$ & 780.027 & 0.0029264\\
    $^{87}$Sr & 86.908884 & $5s^{2}$ $^{1}$S$_{0}-5s5p$ $^{1}$P$^{0}_{1}$ & 460.733 & 0.0049544\\
    $^{93}$Nb & 92.906377 & $4d^{4}$ $(a^{5}$D$)5s$ a$^{6}$D$_{1/2}-4d^{3}$ 5s$(a^{5}$P$)5p$ y$^{6}$P$^{0}_{3/2}$ & 353.530 & 0.0060399 \\
    $^{107}$Ag & 106.905092 & $4d^{10}(^{1}$S$)5s$ $^{2}$S$_{1/2}-4d^{10}(^{1}$S$)5p$ $^{2}$P$^{0}_{3/2}$ & 328.0680 & 0.0056564 \\
    $^{114}$Cd & 113.903357 & $5s^{2}$ $^{2}$S$_{0}-5s5p$ $^{1}$P$^{0}_{1}$ & 228.8022 & 0.0076122 \\
    $^{115}$In & 114.903800 & $5p$ $^{2}$P$^{0}_{1/2}-6s$ $^{2}$S$_{1/2}$ & 410.17504 & 0.0042092 \\
    $^{133}$Cs & 132.905429 & $6s$ $^{2}$S$_{1/2}-6p$ $^{2}$P$^{0}_{3/2}$ & 852.113 & 0.0017517 \\
    $^{153}$Eu & 152.921225 & $4f^{7}6s^{2}$ a$^{8}$S$^{0}_{7/2}-4f^{7}(^{8}$S$^{0})6s6p$ $(^{1}$P$^{0})$ y$^{8}$P$_{5/2}$ & 466.188 & 0.0028011 \\
    $^{173}$Yb & 172.938208 & $4f^{14}(^{1}$S$)6s^{2}$ $^{1}$S$_{0}-4f^{14}(^{1}$S$)6s6p$ $^{1}$P$^{0}_{1}$ & 555.6466 & 0.0020645 \\
    $^{197}$Au & 196.966543 & $5d^{10}6s$ $^{2}$S$_{1/2}-5d^{10}6p$ $^{2}$P$^{0}_{1/2}$ & 267.5954 & 0.0037639 \\
    $^{238}$U & 238.050784 & $5f^{3}(^{4}$I$^{0})6d7s^{2}$ $^{5}$L$^{0}_{6}-5f^{3}6d^{2}7p$ $^{7}$N$_{7}$ & 358.48774 & 0.0023247 \\
    \hline
  \end{tabular}
  \label{table}
\end{table}

Samples of chemical elements are listed in table I with related parameters~\cite{Sansonettia}. Every transition considered here is transition between the ground state $|0\rangle$ and an excited state $|1\rangle$. The wavelengths are corresponding to the resonant frequencies of these transitions, respectively. Speeds of these atoms are clearly different as shown in this table.

\begin{figure}
  \includegraphics[width=14cm]{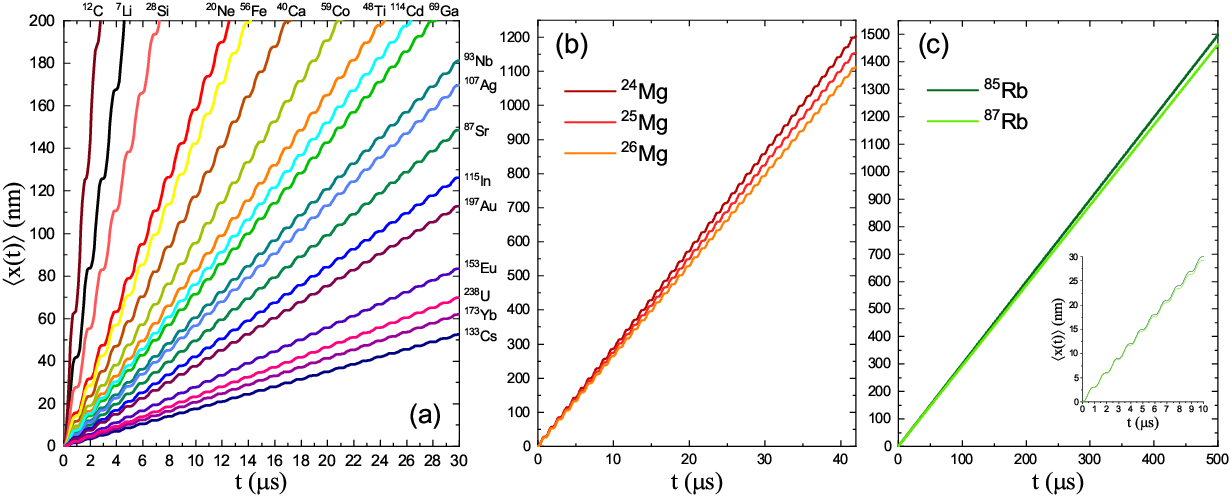}\\
  \caption{(Color on line) (a) Displacements of different atom samples versus time. (b) Displacements of different Magnesium isotopes. (c) Displacements of different Rubidium isotopes.}
  \label{fig3}
\end{figure}

These speed differences give rise to atom deflection as illustrated in Fig.~\ref{fig1}. It is similar to the behavior of magnetic atoms in Stern-Gerlach experiment. The atom displacement in $x$ coordinate can be seen Fig.~\ref{fig3} (a), based on the parameters in Table I. Generally, the coherent walking of light atoms move faster and weight atoms move slower. It is good for separation of these atoms. However, due to there are more selections for the resonant wavelengths, some weight atoms can move faster than light atoms, such as $^{114}$Cd. At the same time, some light atoms may move slower than weight atoms, such as $^{87}$Sr. The fact is benefit to the separation of atoms with similar masses, for example, $^{87}$Rb and $^{87}$Sr, $^{114}$Cd and $^{115}$In. Atoms plotted in Fig.~\ref{fig3} (a) are easily separated each other. The smallest speed difference comes from speed of $^{173}$Yb and speed of $^{238}$U. After $30\mu$s, $^{238}$U atom exceeds $^{173}$Yb atom nearly $10$ nm.

Commonly, isotopes are hardly separated~\cite{Haken}. In the optical material purification, isotopes can be distinguished only relying on theirs masses, since energy structure of isotopes are closed to be the same. Isotopes of light atoms are easier to be separated relatively, such as the Magnesium isotopes shown in Fig.~\ref{fig3} (b). Their distances are nearly $50$ nm at time $42$ $\mu$s. Wight atoms such as Rubidium isotopes $^{85}$Rb and $^{87}$Rb leave each other about $50$ nm after $500$ $\mu$s as illustrated in Fig.~\ref{fig3} (c). The purification ability of the optical method can be further increased by extending coherent time of atoms. This advantage of this method can be seen from Fig.~\ref{fig3} (b) and (c).

\begin{figure}
  \includegraphics[width=8.5cm]{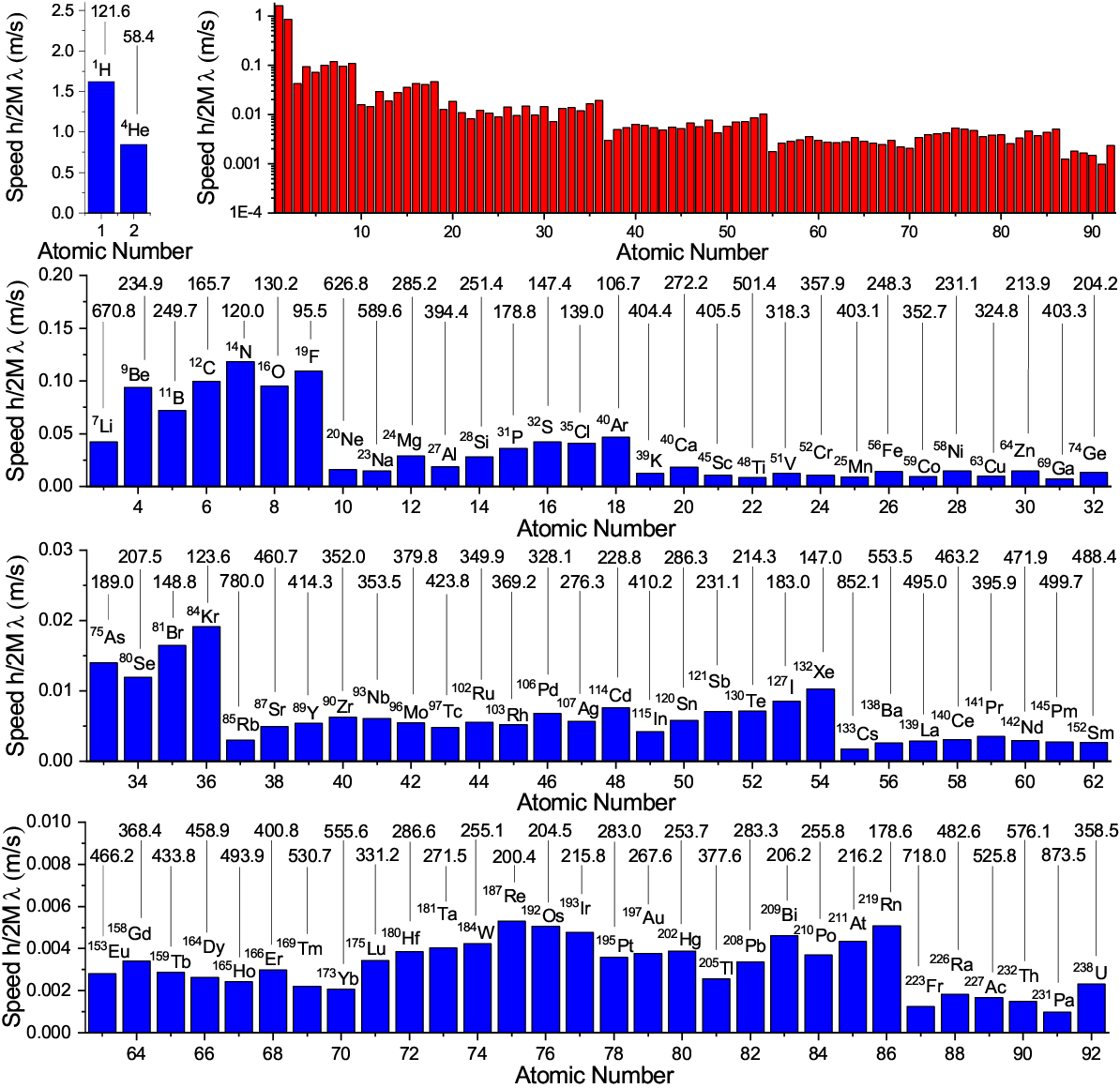}\\
  \caption{Velocity-table of chemical elements for given transition wavelengths. Corresponding wavelength with unit nm is written above each element. The inset is total data bars shown below.}
  \label{fig4}
\end{figure}

A speed table of chemical elements is sketched in Fig.~\ref{fig4} for given transition wavelengths. Speeds of atoms appear several clear ladders, from $^{1}$H to $^{4}$He, from $^{7}$Li to $^{19}$F, from $^{20}$Ne to $^{40}$Ar, from $^{39}$K to $^{84}$Kr, from $^{85}$Rb to $^{132}$Xe, from $^{133}$Cs to $^{219}$Rn and from $^{223}$Fr to $^{238}$U. Obviously, this property is significant for the separation of these chemical elements. The table is not unique that each speed of individual atom can be tuned by selecting other optical transitions.

This optical method should be developed into purification of molecular materials. Indeed, cooling and trapping of molecules has been realized recently by several research groups~\cite{Shuman,Son,Park}. It is known from these reports that masses and transition wavelengths of these cooled molecules are comparable to individual atoms. Furthermore, mass differences between different molecules are generally very large.

In conclusions, we proposed a quantum mechanical method for precise purification of materials based on coherent walking of atoms in light field. A clear advantage of this method is that the separation ability is not only determined on difference of atom masses but also determined on difference of optical transition wavelengths. Therefore, all chemical elements and related isotopes can be separated physically with this configuration in principle. However, weight atoms and weight isotopes are still hard to be separated in fact. Fortunately, as soon as the atom coherent time is long enough, the hardly separable problems should be solved. This method could be mainly used to purify mixed materials which are in the state of fluid or gaseous.

\begin{acknowledgments}
This work was supported by R \& D  Program of Beijing Municipal Education Commission (KM202011232017).
\end{acknowledgments}

\end{document}